\documentstyle[aps,multicol,prl]{revtex}
\input{epsf}

\renewcommand{\r}{{\bf r}}

\newcommand{\dr}{{\bf dr}}
\newcommand{\rp}{{\bf r'}}
\newcommand{\drp}{{\bf dr'}}

\renewcommand{\k}{{\bf k}}

\newcommand{\q}{{\bf q}}
\newcommand{\dk}{{\bf dk}}
\newcommand{\G}{{\bf G}}

\newcommand{\half} {{\frac{1}{2} }}

\begin{document}

\title{Efficient Total Energy Calculations from Self-Energy Models}
\author{Paula S\'anchez-Friera and R.W. Godby}
\address{Department of Physics, University of York, Heslington, York
  YO10 5DD, United Kingdom}
\date{\today}
\maketitle

\begin{abstract}
We propose a new method for calculating total energies of systems of interacting electrons, 
which requires little more computational resources than standard density-functional theories. The total energy is calculated within the framework of many-body perturbation theory by using an efficient model of the self-energy, that nevertheless retains the main features of the exact operator. The method shows promising performance when tested against quantum Monte Carlo results for the linear response of the homogeneous electron gas and structural properties of bulk silicon.\\


\end{abstract}

\begin{multicols}{2}[]

Density-functional theory (DFT) \cite{hk64}
is a powerful method to calculate ground-state properties of
electronic systems. In its standard Kohn-Sham form \cite{ks65}, the system of interacting
electrons is mapped onto a system of non-interacting electrons moving
in an effective local potential. This potential, however, exhibits some non-analyticities in its dependence on the electron density, such as the discontinuity on addition of an extra particle to the system, which is reflected in the band gap problem \cite{godby86} and the failure to correctly describe the response of a macroscopic system to an external electric field \cite{ggg}. This non-analytic behaviour is missing from the standard approximations to the Kohn-Sham potential, 
i.e. the local density (LDA) and generalized gradient
approximations (GGA),
which may explain some of their limitations
when applied to complex systems
such as catalyzed chemical reactions \cite{bird97}.

One way forward is to avoid the need to describe the non-analyticities by incorporating the true non-local nature of the
exchange and correlation. Recently, a new realization of DFT,
the generalized Kohn-Sham scheme (GKS) \cite{seidl96,engel96}, has been proposed, in which the electrons move in an effective \emph{non-local} potential. Among the GKS approaches, the screened-exchange LDA (sX-LDA) \cite{seidl96} appears to give the best performance, describing structural properties with the same accuracy as the LDA but improving on its description of quasiparticle energies. However, this scheme is 
numerically as expensive as a standard Hartree-Fock calculation. 

In this work we propose a new generalized Kohn-Sham scheme, the $\Sigma$-GKS scheme, in which the exchange-correlation potential is intended to mimic the self-energy operator, which 
in many-body perturbation theory (MBPT) exactly describes exchange and
correlation effects and, being truly non-local, is expected to be more amenable to approximation than the Kohn-Sham potential. 

In the framework of MBPT, the total energy of a system of electrons moving in an external potential $V_{ext}$
can be calculated by means of the Galitskii-Migdal formula:
\begin{equation} \label{Eq:gm}
E=\frac{1}{2} \sum_{i,j} \int_{-\infty}^{\mu} d \omega 
\hspace{0.2cm} ( \omega \delta_{i j} 
+ h_{i j} ) A_{j i} (\omega)
\end{equation}
where $i$ and $j$ are the indices of a suitable complete set of one-particle
wavefunctions, $h_{i j}$ are the matrix elements of $h=-\nabla^2/2+V_{ext}$
and $A$ is the spectral function of the Green's
function. Atomic units are used throughout this Letter. 

>From a particular approximation for the self-energy, a model spectral function $A$ can be obtained and the total energy can be calculated via Eq. \ref{Eq:gm}. 
Using the $GW$ approximation \cite{hedin69}, in its self-consistent formulation, total
energies derived from Eq. \ref{Eq:gm} have recently been shown to be in very
good agreement with quantum Monte Carlo data, when applied to 
3D \cite{vonbarth96,garciagonzalez00} and 2D \cite{garciagonzalez00} homogeneous electron gases. However this approximation is computationally too expensive to be in competition with methods such as the LDA and GGA in applications to very large systems. The self-energy model that we propose will allow us to calculate total energies within the same order of numerical cost as the LDA.

We start by modelling the self-energy of 
a homogeneous electron gas (jellium) of density $n_0$. 
For occupied states the energy-dependence of the self-energy
is relatively weak \cite{hedin69}, suggesting the approximation\cite{foot3}:
\begin{equation}
\Sigma(\vert \r - \rp \vert; \epsilon_i) \approx
\Sigma(\vert \r - \rp \vert; \mu)
\end{equation}
where $\mu$ is the chemical potential of the system.
The self-energy for jellium at $\mu$ can be well reproduced by the following model function
\begin{equation} \label{Eq:sig0h}
  \Sigma^o(\r,\rp; n_0) = f(n_0) g(\vert \r - \rp
  \vert; n_0)
\end{equation}
where $f(n_0)=V_{xc}^{LDA}(n_0)$ and 
\begin{equation} \label{Eq:g}
g(\vert \r - \rp \vert;n_0) =  C(n_0) \frac{ e^{-
  \alpha(n_0) \vert \r - \rp \vert }}{\vert \r - \rp \vert }.
\end{equation}
The constant $C(n_0)$ 
ensures that the chemical potential given by $\Sigma^o$ is correct and $1/\alpha(n_0)$ represents the range of the self-energy for a jellium of density $n_0$. 

\begin{figure}[htb] 
\epsfxsize=0.4\columnwidth \epsfbox{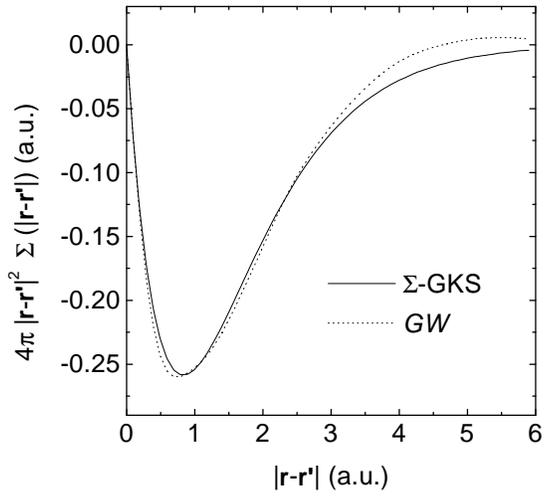}
\caption{Comparison between the model self-energy and the  
  $GW$ data from Ref. [12]
for jellium of $r_s=2$.
\label{fig:smmodel}}
\end{figure}

In Fig. \ref{fig:smmodel} the model function given by Eq. \ref{Eq:sig0h} is compared to the $GW$ data of Ref. \cite{white98}, showing an excellent agreement in the region in which $\Sigma$ has the largest contribution. The 
small discrepancy that appears in the long-range region is unimportant for the total energy, owing to the oscillatory behaviour of the wavefunctions \cite{foot1}. In addition to its suitability for jellium, the function $g$ defined in Eq. \ref{Eq:g} has been shown to have the correct asymptotic decay in semiconducting systems \cite{schindlmayr2000}.

The key idea of this approach is to use a jellium-like self-energy to
describe the exchange and correlation energy of inhomogeneous
systems. This is suggested by the fact that for several semiconductors
the self-energy has been shown \cite{godby88} to be almost spherical
and to have the same range as the self-energy of a jellium with the
corresponding average density. A natural extension of Eq. \ref{Eq:sig0h} to an inhomogeneous system is:
\begin{equation} \label{Eq:sig0}
  \Sigma^o(\r,\rp; [n]) =  \frac {f(n(\r)) + f(n(\rp))}{2} g(\vert \r - \rp \vert; n_0)
\end{equation}
where here $n_0$ is the average density.
This may be likened to a LDA exchange-correlation potential that has been made non-local by including a \emph{spreading function} $g$. 

The self-energy in Eq. \ref{Eq:sig0} is the self-energy of a system of non-interacting electrons, since it is energy independent and real. The method that we propose is equivalent to a mapping between the interacting system and the fictitious non-interacting system represented by $\Sigma^o$. If we choose the basis set $\phi_i$ to be formed by the wavefunctions of the quasiparticles in this fictitious sytem, then the spectral function $A^o$ is simply given by
\begin{equation}
A_{i j}^o(\omega) = \delta_{i j} \delta ( \omega - \epsilon_i^o)
\end{equation} 
where $\epsilon_i^o$ are the quasiparticle energies of the non-interacting system.

Using this spectral function $A^o$ and the Galitskii-Migdal formula (Eq. \ref{Eq:gm}) the total energy of the model system can be obtained:
\begin{eqnarray}
E^o&=&\sum_i^{occ} \left \langle \phi_i \left \vert -\half \nabla ^2 + V_{ext} + \half V_H + \half \Sigma^o \right \vert \phi_i \right \rangle \nonumber \\
&&=T+E_{ext}+E_H+E_{nl}
\end{eqnarray}
where, as usual, $T$ is the kinetic energy, $E_H$ is the Hartree energy, $E_{ext}$ is the external energy and
\begin{eqnarray}
&& E_{nl} = \nonumber \\
&&\frac{1}{2} \sum_i^{occ} \int \dr \hspace{0.2cm} \phi_i^*(\r)
 \int \drp \hspace{0.2cm} \Sigma^o(\r,\rp,[n]) \phi_i(\rp) \label{Eq:enl}.
\end{eqnarray}

As in other generalized Kohn-Sham schemes, a correction term $E_{ss}$ is added so that the total energy given by the model is exact in the homogeneous limit.
This term can be interpreted
in MBPT as the remaining energy associated with
the satellite structure of the spectral function \cite{foot2}.
\begin{equation} \label{Eq:Ess}
E_{ss}[n]=\int \dr \hspace{0.2cm} n(\r) \left (\epsilon_{xc}^{LDA}(n(\r)) - \epsilon_l (n(\r)) \right ) 
\end{equation}
where $\epsilon_l$ is the energy per particle obtained from Eq. \ref{Eq:enl} in the limit of homogeneous densities:
\begin{equation}
  \epsilon_l(n)=\frac{1}{8\pi^3n} \int_{\vert \k \vert < k_F(n)} \dk \hspace{0.2cm} \Sigma(\k)
\end{equation}
where $k_F(n)$ is the Fermi vector of a jellium of density $n$.

The total energy of the interacting system is thus:
\begin{equation} \label{eq:etotgks}
E= T+E_H+E_{ext}+E_{nl}+E_{ss}.
\end{equation}

This functional is minimized with respect to variations in the
one-particle wavefunctions, yielding the following effective hamiltonian:
\begin{eqnarray} \label{Eq:var}
\hat h = - \frac {1}{2} \nabla^2 + V_{ext} + V_{H} +
\left( \frac{1}{2} \Sigma^0 + V_m + V_{ss} \right)
\end{eqnarray}
where $V_{ss}(n(\r))=\delta E_{ss} / \delta n(\r)$
and
\begin{eqnarray} \label{eq:vmdef}
V_m(\r)= \frac {1}{2} f'(n(\r)) \int \drp \hspace{0.2cm} \textrm{Re} \rho(\r,\rp) 
g(\vert \r - \rp \vert),
\end{eqnarray}
in which $\rho$ is the density matrix of the fictitious non-interacting system:
\begin{equation}
\rho(\r,\rp)=\sum_i^{occ} \phi_i^*(\r)\phi_i(\rp)
\end{equation}

In the limiting case of the spreading function $g$ tending to a delta-function, the term $\frac{1}{2} \Sigma^0 + V_m + V_{ss}$ reduces to the conventional LDA exchange-correlation potential, so that the LDA can be considered as a particular case of the more general scheme proposed here.

The computational effort involved in solving the variational equations determined by Eq. \ref{Eq:var} is not significantly larger than in traditional DFT's. In a plane wave basis set, the matrix elements of the non-local potential take the simple form:
\begin{eqnarray}
&&\langle \k+\G \vert \Sigma^0 \vert \k + \G' \rangle \nonumber \\
&& =f(\G-\G')
\frac {g(\vert \k+\G \vert ) + g( \vert \k + \G' \vert )}{2}
\end{eqnarray}
i.e. they are just the product of the LDA exchange-correlation
potential and an analytical function of the moduli of the
vectors $\k+\G$ and $\k+\G'$. The calculation of $\Sigma^0$ scales, as with local potentials, linearly with the number of reciprocal-lattice vectors $n_G$.

The potential $V_m$ defined by Eq. \ref{eq:vmdef} depends on the density matrix of the system rather than the density itself, but it can be efficiently calculated in reciprocal space, where the action of the non-local operator $g$ on the wavefunctions is simply given by the products $g(\G)\phi_i(\G)$. The calculation of $V_m$ also scales linearly with $n_G$. In contrast, in Hartree-Fock or sX-LDA schemes, calculating the exchange-correlation matrix scales with the cube of $n_G$.

\begin{figure}[htb]
\epsfxsize=0.4\columnwidth \epsfbox{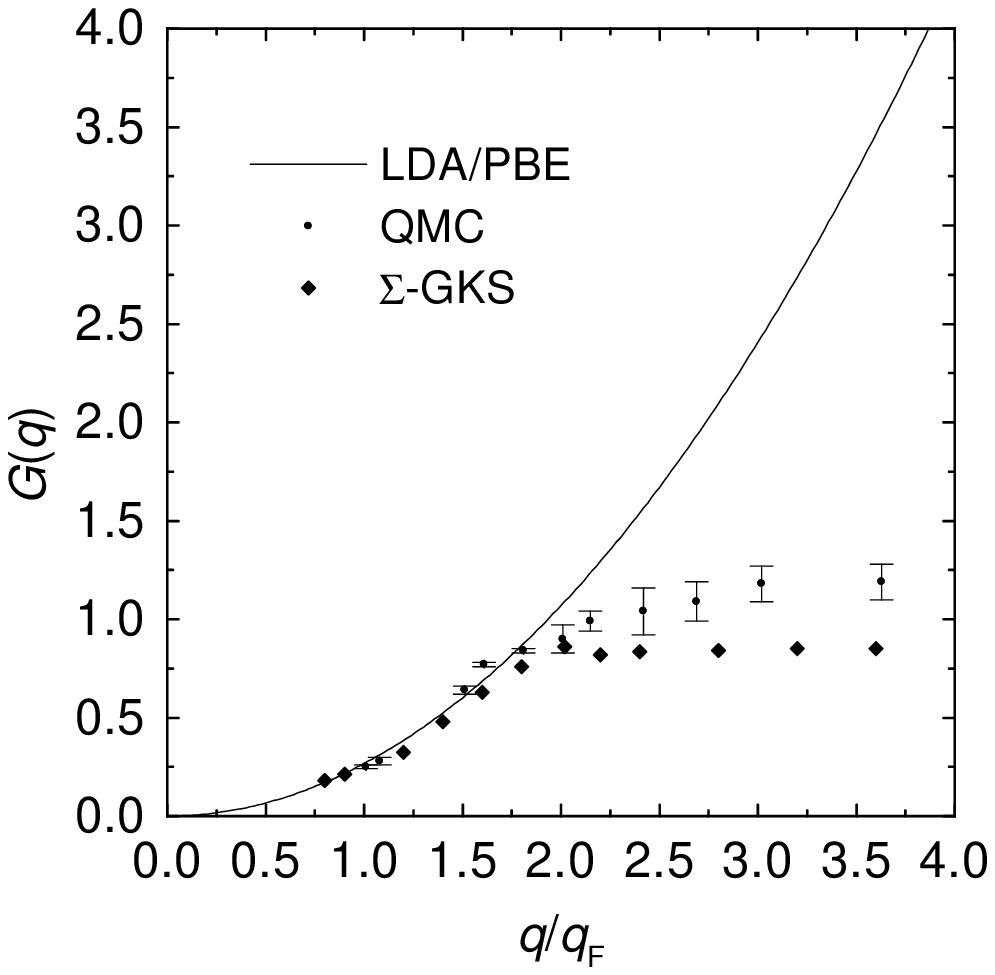}
\caption{Local field factor for the linear response of jellium at $r_s=2$ in the $\Sigma$-GKS scheme compared to
the QMC results [16], the LDA and the GGA of Ref. [17] (PBE) (which coincide). The numerical uncertainties in the $\Sigma$-GKS results are of the order of 0.05 and are not shown for clarity. The new scheme proposed considerably improves the local field factor for large values of $q$ with respect to the LDA and to PBE.
 \label{Fig:lrgks}}
\end{figure}

By construction, the $\Sigma$-GKS is exact for homogeneous densities. In order to test its performance for inhomogeneous systems, we have used it to calculate the linear response of a homogeneous electron gas
to an external perturbation $2\delta V_{ext} cos(\q \cdot \r)$. 
This is a stringent test since it involves the calculation of total energies for a whole family of systems with different values of the perturbation amplitude $\delta V_{ext}$ and the wavevector $q$.
The total energy changes with respect to its unperturbed value by an amount $\chi(\q) \delta V_{ext}^2$, where $\chi(\q)$ is the response function of the system, related to the local field factor $G(\q)$ via:
\begin{equation}
\chi(\q)=\frac{\chi_0(\q) }{1-w(\q)[1-G(\q)]\chi_0(\q)} 
\end{equation}
where $w(\q)=4\pi/q^2$ is the Coulomb potential and $\chi_0(\q)$ is the non-interating Lindhard response function. For a given value of $\q$, the total energy is calculated for several values of $\delta V_{ext}$ and then the set of points ($\delta V_{ext}^2,E$) are fitted to a polynomial, whose first-order coefficient is $\chi(\q)$.
In Fig. \ref{Fig:lrgks}, the local field factor $G(\q)$ calculated within the $\Sigma$-GKS scheme is plotted against the wavevector of the perturbation and
 compared to the quantum Monte Carlo (QMC) results of Moroni {\it et al}
 \cite{moroni95}, the LDA and the GGA. For small values of $q$ all methods agree well with QMC. The large $q$ response, however, is not correctly described with the LDA, nor with the best GGA available, due to Perde, Burke and Ernzerhof (PBE) \cite{pbe}. The $\Sigma$-GKS shows a significant improvement in the local field factor for large values of $q$.

As a further test, the $\Sigma$-GKS scheme was applied to calculate structural properties 
of a typical semiconductor such as silicon. The bulk total energy was calculated using an LDA pseudopotential\cite{fuchs} due to Kerker \cite{kerker} and a plane wave basis set, and compared with the QMC results of Ref. \cite{needs0}, calculated with exactly the same pseudopotential.
The results are shown in Table \ref{tab:si}. Both the $\Sigma$-GKS and the convential DFT's (LDA and GGA) total energies are in good agreement with the QMC results, although $\Sigma$-GKS agrees slightly less well \cite{nextpap}. 
The lattice parameter and bulk modulus are equally well described in the LDA, the GGA and the $\Sigma$-GKS scheme. 

Although describing quasiparticle spectra is not the primary goal of our scheme, we also calculated the quasiparticle energies of Si. The overall tendency is to correctly increase the eigenvalues at the conduction band with respect to the LDA values. The direct band gap at $\Gamma$ is increased from 2.6 eV in the LDA to 3.0 eV in our scheme, whereas the $GW$ and experimental values are 3.4 eV. Valence-band widths in the $\Sigma$-GKS are in general overestimated.

In summary, we have proposed a new method for calculating total energies at the
same time as quasiparticle energies which is just as efficient as a
standard DFT calculation, but that is constructed as a model of the
self-energy, thus describing the exchange and correlation in a more
realistic way than the LDA and GGA. This type of approach holds the prospect of enhanced accuracy of total-energy calculations by avoiding the pathological aspects of the traditional Kohn-Sham exchange-correlation energy functional.

 We gratefully
acknowledge many enlightening discussions with Pablo Garc\'{\i}a-Gonz\'alez and
thank Richard Needs, Paul Kent, Bengt Holm and Ian White for providing useful quantum Monte Carlo and $GW$ data.  This work has been partially funded by the EPSRC and Fundaci\'on Repsol.

\end{multicols}
\begin{table}[htb]

\caption{Total energy of bulk Si within the $\Sigma$-GKS scheme, 
the LDA and GGA (PW91) [22], calculated with a 4$\times$4$\times$4 Monkhorst-Pack $\k$-grid, and QMC [21], all with a cutoff of 15 Ry for the $\G$-vectors. A LDA pseudopotential due to Kerker [20] was used in all four cases, which allows a comparison of the exchange-correlation functional under this particular choice of external potential. The lattice parameter and bulk modulus were also calculated, using a cutoff of 22 Ry for the $\G$-points and a Monkhorst-Pack set of 14 special $\k$-points in the Brillouin zone. The use of an LDA pseudopotential does not justify the comparison of our results with experiment, but nevertheless we present also the experimental values, for completeness. In common with the LDA and PW91, the $\Sigma$-GKS scheme shows generally good agreement with QMC and experiment. Note that quantities shown in parentheses have not been calculated with the pseudopotential of Ref. [20].\label{tab:si}}

\begin{tabular}{cccc}
&$E$ (eV/atom)&$a$ (\AA) & $B$ (MBar)\\
\tableline
$\Sigma$-GKS&-107.65 & 5.39 & 1.008 \\
LDA&-107.90 & 5.39 & 0.967 \\
PW91&-108.17&
(5.46\tablenote{From Ref.\cite{martin97}})&
(0.92$^{\text{a}}$)\\
QMC&-108.3\tablenote{From Ref.\cite{needs0}; estimated finite-size convergence error $\pm 0.05$ eV/atom} & 
(5.45\tablenote{From Ref. \cite{li99}}) & (1.03$^{\text{c}}$)\\
Expt.& -& (5.43\tablenote{Cited in Ref. \cite{li99}}) & (0.992$^{\text{d}}$)
\end{tabular}
\end{table}

\vspace{2cm}

\vspace{2cm}

\begin{thebibliography}{10}

\bibitem{hk64}
P. Hohenberg and W. Kohn, Phys. Rev. {\bf 136},  B864  (1964).

\bibitem{ks65}
W. Kohn and L.J. Sham, Phys. Rev. {\bf 140},  A1122  (1965).

\bibitem{godby86}
J.P. Perdew and M. Levy, Phys. Rev. Lett. {\bf 51}, 1884 (1983); L.J. Sham and M. Schl\"uter, Phys. Rev. Lett. {\bf 51}, 1888 (1983); R.W. Godby, M. Schl\"uter, and L.J. Sham, Phys. Rev. Lett. {\bf 56},  2415  (1986).

\bibitem{ggg}
X. Gonze, P. Ghosez, and R.W. Godby, Phys. Rev. Lett. {\bf 74},  4035  (1995);
P. Ghosez, X. Gonze, and R.W. Godby, Phys. Rev. B {\bf 56},  12811  (1997).

\bibitem{bird97}
D.M. Bird and P.A. Gravil, Surf. Science {\bf 377},  555  (1997).

\bibitem{seidl96}
A. Seidl, A. G\"orling, P. Vogl, J.A. Majewski, and M. Levy, Phys. Rev. B {\bf
  53},  3764  (1996);
G.E. Engel, Phys. Rev. Lett. {\bf 78},  3515  (1997).

\bibitem{engel96}
G.E. Engel and W.E. Pickett, Phys. Rev. B {\bf 54},  8420  (1996).

\bibitem{hedin69}
L. Hedin and S. Lundqvist,  in {\em Solid State Physics}, edited by F.~Seitz
  H.~Ehrenreich and D. Turnbull (Academic Press, New York, 1969), Vol.~23.

\bibitem{vonbarth96}
U. von Barth and B. Holm, Phys. Rev. B {\bf 54},  8411  (1996);
B. Holm and U. von Barth, Phys. Rev. B {\bf 57},  2108  (1998).

\bibitem{garciagonzalez00}
P. Garc\'{\i}a-Gonz\'alez and R.W. Godby (unpublished);
private communication.

\bibitem{foot3} Since the imaginary part of $\Sigma$ at $\mu$ is zero, the model self-energy $\Sigma^0$ is real.

\bibitem{white98}
I. D. White, private communication.

\bibitem{foot1}
The products $\phi_i^*(\r) \Sigma(\r,\rp) \phi_i(\rp)$ cancel out when the
  difference between $\r$ and $\rp$ is large and thus the two wavefunctions are
  not in phase.

\bibitem{schindlmayr2000}
A. Schindlmayr, to be published.

\bibitem{godby88}
R.W. Godby, M. Schl\"uter, and L.J. Sham, Phys. Rev. B {\bf 37},  10159  (1988).

\bibitem{foot2}
A uniform background structure with a certain cutoff $\omega_p$ will yield precisely this
  correction term if $\omega_p$ is chosen using the criterion that the total
  energy of jellium is exact.

\bibitem{moroni95}
S. Moroni, D.~M. Ceperley, and G. Senatore, Phys. Rev. Lett. {\bf 75},  689
  (1995).

\bibitem{pbe}
J.P. Perdew, K. Burke and M. Ernzerhof, Phys. Rev. Lett {\bf 77}, 3865 (1996).

\bibitem{fuchs}
It should be noted that the use of an LDA pseudopotential tends to reproduce LDA structural properties when used with any other exchange-correlation functional: see M. Fuchs, M. Bockstedte, E. Pehlke and M. Scheffler, Phys. Rev. B {\bf 57}, 2134 (1998). This does not affect, however, the comparison between theoretical methods.

\bibitem{kerker}
G. Kerker, J. Phys. C {\bf 13}, L189 (1980).

\bibitem{needs0}
P.R.C. Kent, R.J. Needs and C. Rajagopal, Phys. Rev. B {\bf 59}, 12344 (1999); private communication.

\bibitem{pw91}
J.P. Perdew, J.A. Chevary, S.H. Vosko, K.A. Jackson, M.R. Pederson,
D.J. Singh, and C. Fiolhais, Phys. Rev. B {\bf 46}, 6671 (1992).

\bibitem{martin97}
I.-H. Lee and R.M. Martin, Phys. Rev. B {\bf 56}, 7197 (1997).

\bibitem{li99}
X.-P. Li, D.M. Ceperley and R.M. Martin, Phys. Rev. B {\bf 44}, 10929 (1991).


\bibitem{nextpap}
An improved $\Sigma$-GKS total energy functional can be built through a simple spectral function model for the energy-dependence of $\Sigma$. This reduces the correction term of Eq. \ref{Eq:Ess} by two orders of magnitude in jellium at $r_s=2$ and will the subject of a future publication.

\end{thebibliography}
\end{document}